*Working Paper*

# Is the price right? Reconceptualizing price and income elasticity to anticipate price perception issues

**Shawn Berry, DBA[1*]**

**February 6, 2024**

[1]William Howard Taft University, Lakewood CO, USA

*Correspondence: shawn.berry.8826@taftu.edu

**Abstract** Price perception by consumers represents a challenge to the ability of a business to correctly and profitably price and sell their products or services in a given market and any new target market. Complicating the perception of prices is the dynamics of price and income elasticity, both of which are key for estimating demand. This article proposes a novel metric that conceptualizes elasticity as a means of quantifying the potential for price perception problems in a market using an elegant and non-utility based identity. Elasticity studies from 1990 to 2023 (n=30) were sampled to evaluate the relationship between price and income elasticity for various consumer commodities using the identity. The results suggest that, given known price and income elasticity values, a business can anticipate pricing perception problems in advance and address the potential for damaging distortion of their value proposition. Further, the business can use this insight to correctly choose a strategic pricing approach.

---

## 1. Introduction

The relationship of prices to price elasticity and income elasticity in the realm of consumer behavior is important. Indeed, a competitive advantage through strategic pricing can potentially be obtained if there is a deep understanding of the price and income sensitivities of consumers in a given market (Thompson and Coe, 1997). Kaczala (2023) observes that “knowledge of income elasticity of demand for individual products is necessary to forecast the volume and structure of demand” (p.83). In their examination of how price elasticity at the store level is connected to consumer attributes, Hoch et al. (1995) found “11 demographic and competitive variables explain on average 67% of the variation in price response” (p.17), adding “that the consumer demographic variables are much more influential than competitive variables” (Hoch et al., 1995, p.17). The observation by Hoch et al. (1995) implies a connection to income elasticity as a determinant of how prices are perceived by consumers. Price perception is an important aspect of consumer behavior, particularly the role of reference prices (Meng, Suzanne, and Clark, 2008). Work by Eisenhauer and Principe (2009) on the topic of reference price knowledge of consumers in the context of price elasticity found “that price knowledge is positively related to price sensitivity” (p.14). These observations suggest that both price elasticity and income elasticity are each fundamental for understanding the sensitivity of consumers in the marketplace.

Although many studies on price elasticity and income elasticity have been published over several decades by many scholars, the literature presents some limitations to their usefulness over time. First, a cursory examination of the literature suggests that the generalization of many studies to consumer commodities on a global basis is lacking because of the narrow focus of articles to the consumption of electricity, alcohol, or cigarettes in a specific nation, for example. Second, a cursory examination of the literature also indicates that studies do not seem to consistently report both price and income elasticity together, and usually tend to report only price elasticity alone. This inconsistency does not allow the evaluation of reliability or comparison among other studies over time, particularly for commodity groups or products. Gallet and Doucouliagos (2014) reported that "there is much disparity in the estimates of income elasticity of air travel across the literature" (p.141). Third, with the passage of time and alterations in the business environment, the validity of previous elasticity estimates has not kept pace with alterations in the business and market environments and approaches used by academics (Bijmolt, Van Heerde, and Pieters, 2005). Finally, Sabatelli (2016) identifies a gap in knowledge, stating that "to-date no published study has investigated the relationship between income elasticity and price elasticity of demand" (p.12).

In this short article, a non-utility based identity is presented to generalize price perception error by consumers using the relative price and income elasticities of a product or service in a given market. Specifically, the goal of this article is to extend prior work by others, and help to fill the gap in knowledge to the extent of providing an application of both price and income elasticity to consumer pricing. In particular, the identity will reconceptualize elasticity in a novel way that can be applied by a firm to address potential pricing conflicts in any market, and identify competitive advantages for the firm at a consumer level. This insight is important to the creation and monitoring of pricing strategy, and critical to the firm gaining market share, profit, and consequently, a competitive advantage. Using known price and income elasticity data estimates from the literature, the study will present the use of the identity to identify price perception issues. Finally, the study will conclude with a discussion of the results and directions for future research,

## 2. Materials and Methods

The sample data used in this study has been drawn from journal articles from 1990 to 2023 (n=30) that reported both price elasticity and income elasticity estimates for various consumer products and services. Articles that were focused on price elasticity of public utilities, such as water and electricity, for example, or only price elasticity alone were excluded from the sample. As a result, the sample data collected from journal articles represents a global mélange of observed price and income elasticities over time.

The specification for the identity to describe the price perception issue for a consumer in the context of elasticity is given below in Eq. (1) as:

$$\frac{Pa - Pr}{Pa} = \frac{\eta p}{\eta i} - 1 \tag{1}$$

where Pa is the actual price of the product or service, Pr is the reference price for the product or service, $\eta p$ and $\eta i$ are the corresponding price and income elasticities, respectively, as proposed by Eisenhauer and Principe (2009, p.12) in their work on pricing knowledge. If the left side of Eq. 1 is positive (Pa > Pr), the consumer has underestimated the price, and if the left side is negative, the consumer has overestimated ($Pa < Pr$) (Eisenhauer and Principe, 2009, p.12). On the right side of Eq. 1, this paper proposes that the consumer will have a price perception error of zero when both the price elasticity and the income elasticity are the same, given that subtracting 1 from this ratio when the ratio is equal to 1 will result in zero. This implies that the price and income elasticity are such that the firm has correctly determined what the price and income response of the consumer will be in a given market or for a given commodity or service. This paper proposes that while a negative ratio will imply that consumers may tend to overestimate the actual price in the market, a positive ratio will imply that consumers may tend to underestimate the actual price in the market. The equation is elegant in that it can be rearranged to solve for any variable, given known values. The firm can test values of reference prices used by consumers, given known price and income elasticity values, to compute an actual price level that will not incur price perception error.

.

## 3. Results

The sample data was analyzed, and the proposed identity was applied.

### *3.1. Descriptive statistics*

Table 1 summarizes the descriptive statistics for the sample data. The mean and standard deviation for price elasticity are reasonably close to 0 and 1, respectively. The Shapiro-Wilks statistics for each of price and income elasticity suggest that the data are drawn from a normally distributed population. The standard errors for both price and income elasticity are comparable. The mean for income elasticity is much higher than the mean for price elasticity, and the standard deviation for income elasticity is less than that for price elasticity. The range between the maximum and minimums of price elasticity are much wider than that for income elasticity. Although the Shapiro-Wilks statistic for the ratio of price elasticity and income elasticity is not significant, the histogram in Figure 1 suggests that the values approximate a normal distribution.

**Table 1.** Descriptive statistics.

| | ηp | ηi | ηp / ηi |
|---|---|---|---|
| Mean | 0.02 | 0.58 | -0.89 |
| Median | -0.33 | 0.45 | -0.68 |
| Minimum | -1.56 | -0.9 | -8.56 |
| Maximum | 2.12 | 2.62 | 7.36 |
| S.E. | 0.19 | 0.14 | 0.65 |
| S.D. | 1.04 | 0.79 | 2.75 |
| Shapiro-Wilks (p-value) | 0.933 (0.06) | 0.967 (0.467) | 0.919 (0.03) |
| n | 30 | 30 | 30 |

**Source:** data analysis

**Figure 1:** Histogram of ratio of price elasticity and income elasticity

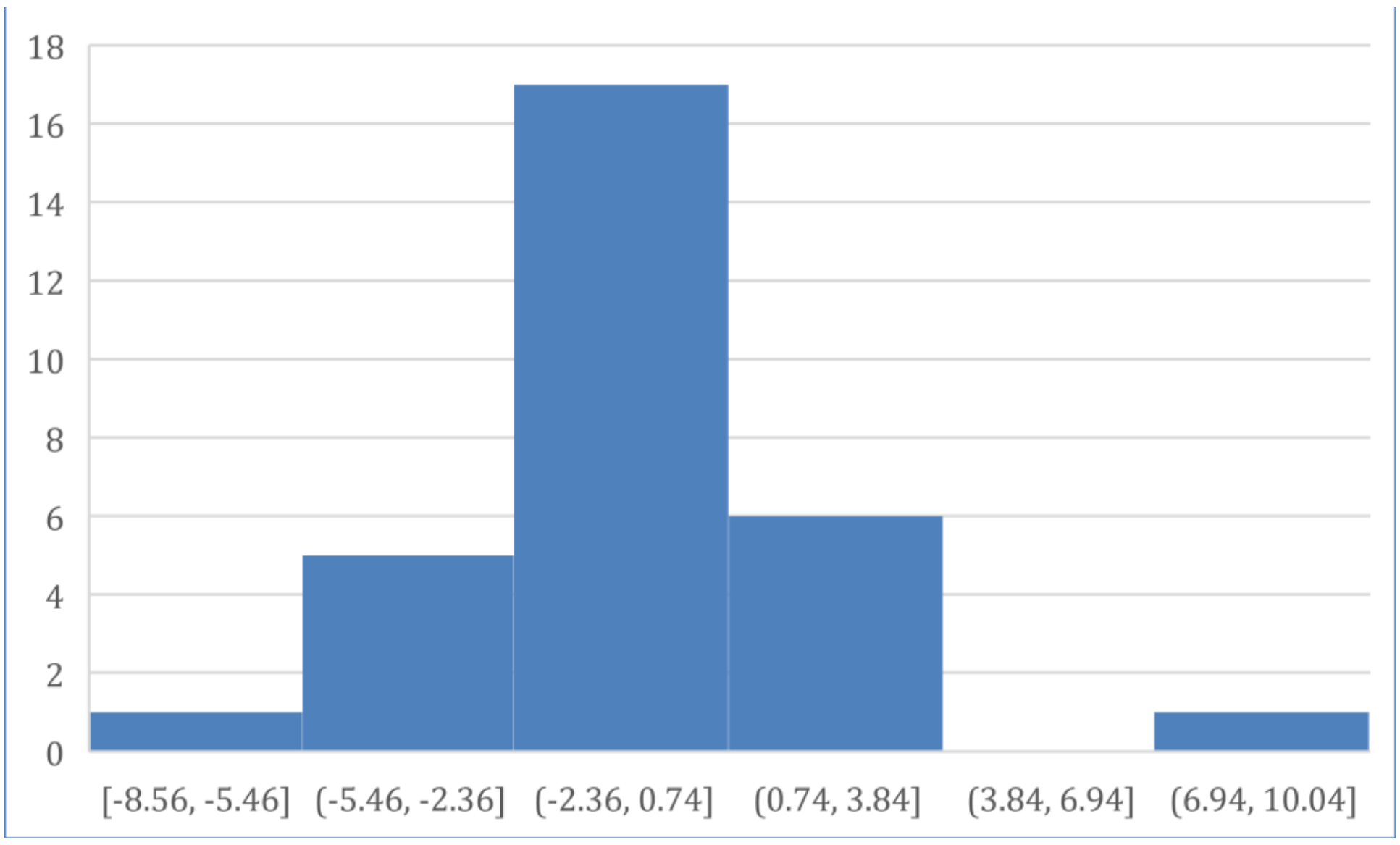

**Source:** data analysis

### 3.2 *Sample data*

The sample data is presented in Table 2 below, and it is sorted in order according to the column containing the consumer price perception error, represented by the column (ηp/ηi)-1.

Table 2 generally illustrates that commodities or services with large price elasticities and small income elasticities lead to large consumer price perception errors where consumers would have used an overpriced reference price as compared with the actual price. Examples of these errors at the extreme end of the spectrum appear to be mainly organic vegetables, frozen desserts, and real estate services. At the other end of the spectrum, where values are positive, are US dairy products and non-organic onions. Consumer price perception error values that were close to zero or very small, indicating that price and income elasticity were very similar, were mainly Indonesian meat, dairy, and fish products, and to a lesser extent, US sugar and US fish products.

**Table 2.** Price and income elasticities of products and services, elasticity ratios, price perception errors, and sources.

| Commodity/Service | ηp | ηi | ηp/ηi | (ηp/ηi)-1 | Source |
|---|---|---|---|---|---|
| Non-organic potatoes | 1.54 | -0.18 | -8.56 | -9.56 | Trost, 1999 |
| Organic onions | -1.56 | 0.32 | -4.88 | -5.88 | Trost, 1999 |
| Frozen dessert products | -0.356 | 0.08 | -4.45 | -5.45 | Kaiser & Forker, 1993 |
| Oranges in South Africa | -1.55 | 0.407 | -3.81 | -4.81 | Hayward-Butt & Ortmann, 1994 |
| Organic vegetables in Taiwan | -0.152 | 0.04 | -3.80 | -4.80 | Huang-Zheng and Lin, 2011 |
| Real estate services | -1.20 | 0.40 | -3.00 | -4.00 | Bates & Santerre, 2016 |
| Poultry, USA | -1.313 | 0.659 | -1.99 | -2.99 | Young, 1990 |
| Pork, USA | -0.854 | 0.507 | -1.68 | -2.68 | Young, 1990 |
| Theatre tickets in Switzerland | 0.3 | -0.2 | -1.50 | -2.50 | Zieba, 2016 |
| Theatre tickets in Austria | 0.7 | -0.5 | -1.40 | -2.40 | Zieba, 2016 |
| Vegetables, USA | -0.421 | 0.313 | -1.35 | -2.35 | Young, 1990 |
| Eggs in South Africa | -0.55 | 0.41 | -1.34 | -2.34 | Cleasby and Ortmann, 1991 |
| Standard cheese in Norway | -1.009 | 1.121 | -0.90 | -1.90 | Sooriyakumar, 2003 |
| Tea in Iran | -0.42 | 0.53 | -0.79 | -1.79 | Fallah Alipour, Kavoosi Kalashami and Ahmedzedah, 2019 |
| Jam in Peshawar | -0.46 | 0.64 | -0.72 | -1.72 | Khanum et al., 2007 |
| Cigarettes in Bangladesh | 0.39 | -0.62 | -0.63 | -1.63 | Ahmed et al., 2022 |
| Beer | -0.3 | 0.5 | -0.60 | -1.60 | Nelson, 2013 |
| Public bus transport | -0.59 | 1.05 | -0.56 | -1.56 | Holmgren, 2007 |
| Spirits | -0.55 | 1 | -0.55 | -1.55 | Nelson, 2013 |
| Wine | -0.45 | 1 | -0.45 | -1.45 | Nelson, 2013 |
| Durable goods | -0.49 | 1.35 | -0.36 | -1.36 | Wong & McDermott, 1990 |
| Sugar, USA | -0.294 | -0.898 | 0.33 | -0.67 | Young, 1990 |
| Milk in Indonesia | 1.32 | 1.84 | 0.72 | -0.28 | Forgenie, Khoiriyah and Elbaar, 2023 |
| Beef in Indonesia | 1.71 | 2.2 | 0.78 | -0.22 | Forgenie, Khoiriyah and Elbaar, 2023 |
| Fish, USA | 2.124 | 2.616 | 0.81 | -0.19 | Young, 1990 |
| Poultry in Indonesia | 1.39 | 1.44 | 0.97 | -0.03 | Forgenie, Khoiriyah and Elbaar, 2023 |
| Fish in Indonesia | 1.11 | 1.07 | 1.04 | 0.04 | Forgenie, Khoiriyah and Elbaar, 2023 |
| Cereal, USA | -0.045 | -0.029 | 1.55 | 0.55 | Young, 1990 |
| Dairy, USA | 0.645 | 0.21 | 3.07 | 2.07 | Young, 1990 |
| Non-organic onions | 1.84 | 0.25 | 7.36 | 6.36 | Trost, 1999 |

### *3.3. Regression analysis: Price elasticity versus income elasticity*

The sample data was analyzed using regression analysis. A scatter plot of the data appears below in Figure 2, and suggests a weak linear relationship between price elasticity and income elasticity. Sabatelli (2016) indicates that "a quantitative (parabolic) functional relationship exists between the income elasticity of demand…and the uncompensated own price elasticity of demand" (p.6). The scatter plot suggests the possibility of such a parabolic relationship between the variables.

**Figure 2.** Scatterplot of price and income elasticity of products and services, 1990-2023

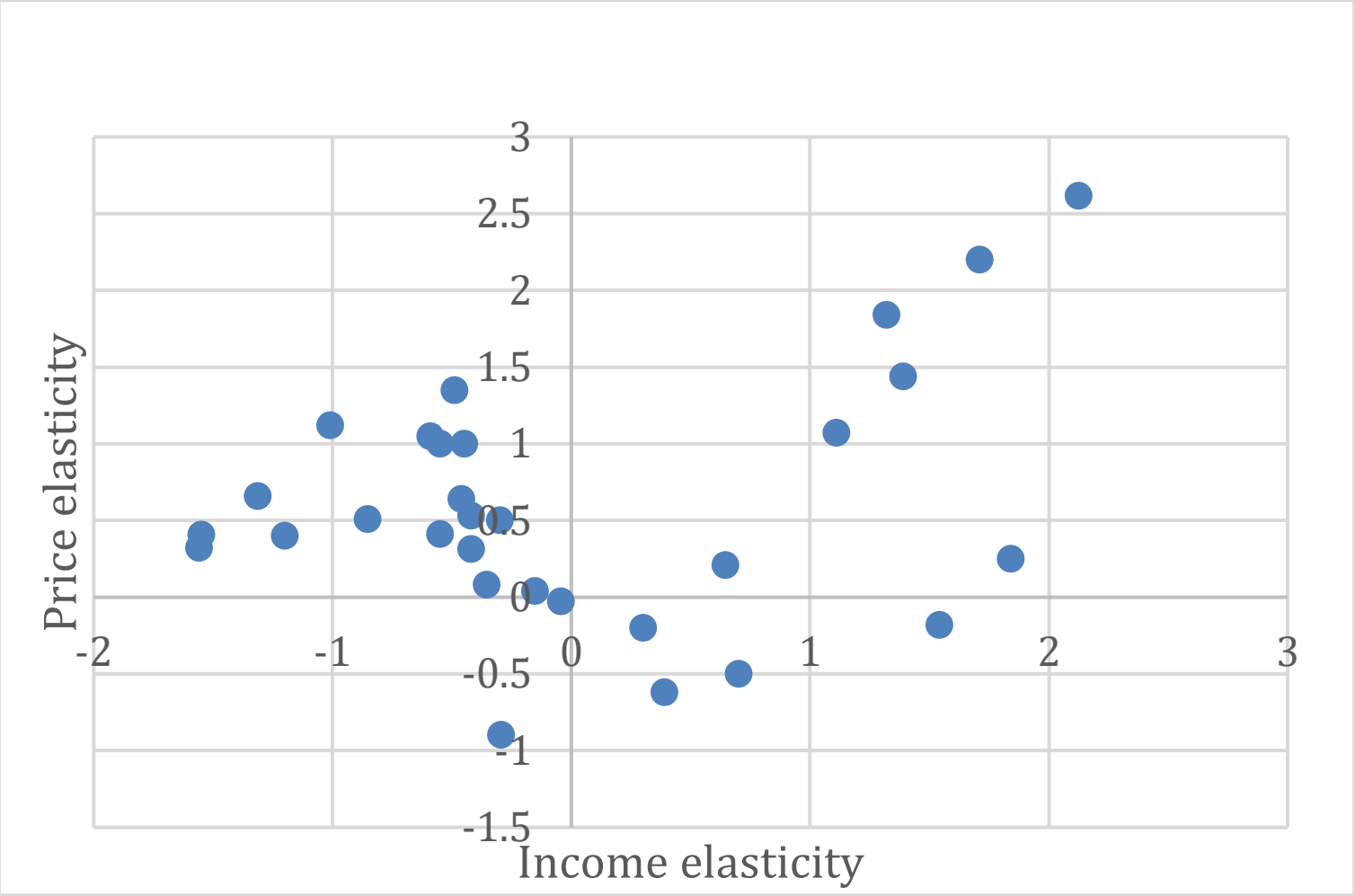

**Source:** data analysis

Using the insight from Sabatelli (2016), a quadratic regression analysis was performed using price elasticity as the dependent variable, and income elasticity as the independent variable. The results of the regression appear in Table 3 below. The independent variable was significant at the 5% level for the non-squared term, and at 1% for the squared term. Although the r-squared is 0.3687 ($F$=7.884, $p$=0.002, $DF$ = 27), the results are much stronger than the linear regression of the data as shown in Table 4 ($r^2$= 0.091, $F$=2.803, $p$=0.105, $DF$=28), and confirming Sabatelli's (2016) observation that the relationship between price elasticity and income elasticity is indeed parabolic.

**Table 3.** Quadratic regression results. Dependent variable: price elasticity

| Variable | Estimate | S.E. | t-value | PR (>\|t\|) | Significance |
|---|---|---|---|---|---|
| Intercept | -0.2131 | 0.195 | -1.092 | 0.285 | |
| ηi | -0.6249 | 0.358 | -1.747 | 0.092 | (*) |
| $\eta i^2$ | 0.6280 | 0.182 | 3.446 | 0.002 | (**) |

Signif. codes : 0 '***' 0.001 '**' 0.01 '*' 0.05 '.' 0.1 ' ' 1

**Source :** data analysis

**Table 4.** Linear regression results. Dependent variable: price elasticity

| Variable | Estimate | S.E. | t-value | PR (>\|t\|) | Significance |
|---|---|---|---|---|---|
| Intercept | -0.214 | 0.230 | -0.932 | 0.359 | |
| ηi | 0.396 | 0.237 | 1.674 | 0.105 | |

Signif. codes : 0 '***' 0.001 '**' 0.01 '*' 0.05 0.1 ' ' 1 '.'

### *3.4. Regression analysis: Ratio* **ηp / ηi**

The relationship of the ratio of price elasticity to income elasticity was analyzed through log transforming the dependent and independent variables. Linear regression was then employed using the log transformed ratio against the log transformed price elasticity and income elasticity. Tables 5 and 6 illustrate the results of the regression of the ratio versus each independent variable separately. In Table 5, the regression explained 41.53% of the variance of the ratio by income elasticity ($F$=19.89, $DF$=28, $p$<0.001). However, in Table 6, only 5.27% of the variance in the ratio was explained by price elasticity ($F$=1.557, $DF$=28, $p$=0.222).

**Table 5.** Regression results. Dependent variable: ηp / ηi

| Variable | Estimate | S.E. | t-value | PR (>\|t\|) | Significance |
|---|---|---|---|---|---|
| Intercept | -0.095 | 0.152 | -0.622 | 0.539 | |
| log(ηi) | -0.531 | 0.119 | -4.460 | 0.0001 | (**) |

Signif. codes : 0 '***' 0.001 '**' 0.01 '*' 0.05 '.' 0.1 ' ' 1

**Table 6.** Regression results. Dependent variable: ηp / ηi

| Variable | Estimate | S.E. | t-value | PR (>\|t\|) | Significance |
|---|---|---|---|---|---|
| Intercept | 0.389 | 0.181 | 2.156 | 0.039 | (*) |
| log(ηp) | 0.241 | 0.193 | 1.248 | 0.222 | |

Signif. codes : 0 '***' 0.001 '**' 0.01 '*' 0.05 '.' 0.1 ' ' 1

## 4. Discussion

.

In broad terms, the use of the identity on the sample data suggests that the relationship between price elasticity and income elasticity has important implications for global marketers of various consumer products and services. First, given the global nature of the sample data, the use of the identity allows marketers to evaluate the potential for price perception problems in various countries. In this instance, the implication for marketers is that the identity allows for easy comparison among countries that may be markets for a given product or commodity, and identification of potential strategic differentiation opportunities in a region. Second, given the finding that organic products appear to have a very high negative ratio relative to non-organic products in the sample data, for example, the implication for marketers is that use of the identity can help strategically differentiate those products and commodities that are generic from those that are specialized. Since price perception problems may suggest a disconnect of value to price on the part of consumers, the insight from the identity can suggest a refocus to the proposed value proposition, given a problematic ratio. Finally, the relativity of price elasticity to income elasticity as given by the ratio is an important instrument for marketers to evaluate not just their own products but also competing products, given known competitive information in a particular market. In the context of price perception, and in a strategic context, marketers can use the identity to evaluate the performance of their own product relative to a competing substitute, inferior, or superior product in the market. Using known consumer reference prices and actual prices from market research, the rearrangement of the identity allows marketers to solve and estimate for any unknown information about their own or a competitor's products.

The statistical analysis of the sample data illustrates several key findings about the relationship between price elasticity and income elasticity, and their implications for anticipating price perception problems. First, the quadratic regression results of price elasticity to income elasticity confirm the observation by Sabatelli (2016) that price elasticity and income elasticity are interrelated, and in a parabolic manner. Furthermore, this study found the relationship to be statistically significant. Second, the log-transformed ratio of price elasticity to income elasticity was found to be statistically significant when it was regressed against the log-transformed values of income elasticity but not for log-transformed price elasticity. Furthermore, since more of the variance of the log-transformed ratio of price elasticity to income elasticity was explained by log-transformed income elasticity than with the log-transformed price elasticity, this observation implies that income elasticity is a more important indicator of consumer price sensitivity. This result echoes the observation by Hoch et al. (1995) that suggested that demographic indicators play a large role in explaining consumer behavior.

The application of the price perception identity to the sample data suggests that the ratio of price elasticity to income elasticity provides useful insights into the interplay of elasticity with the potential for price perception issues for various commodities and services. Furthermore, given the global nature of the sample data, this novel analysis illustrates the importance of understanding the ratio of price elasticity and income elasticity for products in different global settings, particularly if a firm intends to enter a new non-domestic market or product category. More importantly, the identity allows marketers to understand the potential for pricing to not be in alignment with consumer expectations, given the ratio of elasticities. Finally, the identity can be rearranged to allow marketers to test price levels and understand the levels of willingness to pay of consumers by way of their expectations or reference prices, given a known price elasticity and income elasticities of a given subset of consumers. In the context of strategic pricing, particularly where a firm wishes to venture into different global regions or even different domestic regions, the sample data clearly shows that price elasticity and income elasticity must be viewed in tandem to align product prices with regional differences in willingness and ability to pay. The alignment of price elasticity and income elasticity appears to be a critical factor in avoiding price perception issues.

## 5. Directions for future research

This paper illustrates the value of using both price elasticity and income elasticity in the context of generalizing price perception error potential for products in any given market. Market researchers should attempt to identify the reference prices and income elasticities of consumers to identify more granular market segments. To make price elasticity more useful, and to address shortcomings noted in the literature to avoid disparities, reported results must also reflect the income elasticity to understand this aspect of consumer behavior. More research is required to evaluate the degree of price perception error in practice according to the varying income elasticity and price sensitivity levels of consumers. In this regard, marketers can better understand consumer behavior on a more granular level.

## 6. Limitations

Although this paper presents an approach to measuring the potential for price perception problems, the limitations herein must be acknowledged. First, the selection of literature for the sample excluded utilities and other goods that were public, such as electricity and drinking water, for example. Second, since the use of the terms compensated and uncompensated price elasticity was inconsistent in older literature, the price elasticity that was reported in any given paper was recorded without regard to compensation. Finally, since this paper sought to examine the elasticities of consumer goods and services, few studies were found in the literature that reported both price and income elasticity. As a result, the sample data mainly represents the available information about price and income elasticities for food and other consumer products on a global basis rather than the United States itself.

## 7. Patents

There are no patents resulting from the work reported in this manuscript.

## 8. Funding

This research received no external funding.

## 9. Conflicts of Interest

The authors declare no conflict of interest.

## 10. Declaration of generative AI in scientific writing

The author declares that generative AI tools were not used in the writing or research of this article.